\documentclass[alpha-refsdoi]{wiley-article}

\usepackage{color}
\usepackage{ulem}
\usepackage{amsmath}

\usepackage{siunitx}
\usepackage{parskip}
\usepackage{times}
\usepackage{caption}
\usepackage{subcaption}
\usepackage{comment}
\usepackage{xurl}
\usepackage{hyperref}
\usepackage{xr}
\usepackage{cleveref}
\usepackage{lineno}
\papertype{Original Article}
\paperfield{Journal Section}

\title{Impact of Background Conditions on the Structure and Propagation of the Boreal Summer Quasi-Biweekly Oscillation}

\abbrevs{QBWO, Quasi-biweekly Oscillation; OLR, Outgoing Longwave Radiation; JJAS, June-July-August-September.}

\author[1\authfn{1}\authfn{2}]{Shubhrangshu Biswas}
\author[1,2\authfn{1}\authfn{2}]{Jai Sukhatme}
\author[1,3\authfn{1}\authfn{2}]{Bishakhdatta Gayen}

\contrib[\authfn{1}]{Equally contributing authors.}

\affil[1]{Centre for Atmospheric and Oceanic Sciences, Indian Institute of Science, Bengaluru, Karnataka-560012, India}
\affil[2]{Divecha Centre for Climate Change, Indian Institute of Science, Bengaluru, Karnataka-560012, India}
\affil[3]{Mechanical Engineering Department, University of Melbourne, Melbourne, VIC-3010, Australia}

\corraddress{Shubhrangshu Biswas, Centre for Atmospheric and Oceanic Sciences, Indian Institute of Science, Bengaluru, Karnataka-560012, India}
\corremail{shubhrangshu@iisc.ac.in}

\presentadd[\authfn{2}]{Centre for Atmospheric and Oceanic Sciences, Indian Institute of Science, Bengaluru, Karnataka-560012, India}

\fundinginfo{Shubhrangshu Biswas acknowledges Prime Minister's Research Fellowship, Indian Institute of Science and Grantham Fellowship for supporting his Ph.D.}

\runningauthor{Biswas et al.}

\begin{document}
\maketitle

\begin{abstract}

We examine the westward-propagating quasi-biweekly oscillation (QBWO) during boreal summer, with a focus on how background moisture and winds shape its structure and propagation. In dry regions, convection lags the circulation by nearly a quarter cycle, whereas in very moist regions it becomes nearly in-phase and extends across the QBWO gyre. As the background moistens, moisture anomalies increasingly align with the QBWO circulation. Despite differences in environmental moisture and wind conditions, several structural features remain robust: outgoing longwave radiation and moisture anomalies stay collocated, moisture and pressure-velocity anomalies remain vertically upright, and the filtered winds retain a first-baroclinic mode structure. A vorticity budget shows that, although the planetary vorticity-gradient term is important, both planetary stretching and horizontal advection are needed to explain the vorticity tendency—and their relative importance shifts with the moisture regime. In dry and moderately moist regions with easterly mean flow, mean winds primarily advect vorticity anomalies. In contrast, in very moist regions with westerly flow, anomalous winds instead advect the background vorticity. An analogous transition occurs in the moisture budget: in dry and moderately moist environments, zonal mean flow advection dominates, but in very moist regions, strong background moisture gradients allow eddy advection of the mean moisture field to become the leading term. In the moist regime, vertical advection, precipitation, and evaporation also contribute substantially to the moisture tendency. Overall, the QBWO behaves like a mean-flow-driven linear mode in dry and moderately moist regions with easterly background winds, but shifts toward a regime dominated by eddy advection of background vorticity and moisture in very moist regions characterized by westerly flow.

\keywords{Equatorial Rossby waves; QBWO; tropical meteorology; vorticity; moisture}
\end{abstract}
\section{Introduction} 

In the vast continuum of multiscale interactions that govern our atmosphere—spanning processes from hourly to yearly timescales, and from microscopic turbulence to planetary waves—each scale leaves an indelible imprint on weather and climate \citep{klein2010scale}. Among this spectrum, we focus on the intriguing band of quasi-biweekly oscillations (QBWO), whose temporal rhythm of 10–25 days bridges synoptic and intraseasonal variability, and has been observed across much of the tropical belt \citep{kikuchi2009global}. These oscillations exert a profound influence on regional weather patterns, modulating rainfall and synoptic events \citep[for example,][]{goswami-rain}, heat waves \citep{han-heat}, and at times setting the stage for extreme events such as cyclogenesis through their rotational circulation and convective organization \citep{cyclone,ling}. While the most visible and destructive atmospheric phenomena often occur at smaller scales, their genesis is frequently linked to these larger-scale precursors that govern the environment’s propensity for such extremes.

The QBWO was first identified in the context of the Indian summer monsoon \citep{keshavamurty1972vertical, krishnamurti1976} as a coherent large-scale mode. Its structure, energetics, and impacts have been extensively examined in diverse regions, including the northwestern Pacific, the South China Sea, and across South and Southeast Asia \citep{krishnamurti198010, chen199310, fukutomi199910, chatterjee2004structure, chen2010characteristics, jia2013impact, chen2024}. Notably, recent studies have also revealed its imprint on upper-ocean variability, influencing sea surface temperature and salinity on similar timescales \citep{sree2020quasi}. It is important to note that the term QBWO does not specify eastward or westward propagation, and depending on the geographic location and season, movement in either direction is possible \citep{kikuchi2009global}—including significant poleward propagation in both hemispheres \citep{wang2017quasi, yangSH, sambrita}. Much of the aforementioned work, however, has focused on westward-propagating systems, which are thought to be manifestations of convectively coupled Equatorial Rossby (ER) waves \citep[CCERs;][]{kiladis2009convectively} modified by the prevailing regional mean flow \citep{chatterjee2004structure, kikuchi2009global, chen2010characteristics}.

The ER wave is a large-scale atmospheric disturbance that forms naturally near the equator due to the variation of Earth’s rotation with latitude (the so-called $\beta$ effect) and is formally one of the solutions to the dry shallow-water equations on an equatorial $\beta$-plane \citep{matsuno1966quasi}. This wave propagates westward on a background state of rest and exhibits alternating cyclonic–anticyclonic gyres in quadrature with regions of convergence and divergence that are symmetric about the equator \citep{matsuno1966quasi, gill1980some}. In terms of the observed variability of the tropics, ERs are usually identified by isolating a specific region in the wavenumber–frequency domain that adheres to the dispersion curves of Rossby waves \citep{wk1999}—though there are other methods, as recently reviewed by \cite{Kreview}. The reconstructed physical fields corresponding to this window in spectral space (i.e., the selected range of wavenumber–frequency combinations consistent with Rossby wave dynamics) show reasonable agreement with dry theory, but differ in significant ways, including a reduced equivalent depth (a lower effective vertical scale of motion compared to the dry shallow-water solution) and a shift in the location of moist anomalies with respect to the gyres \citep{kw1995, WKW, nakamura2022convective,matthews2025vorticity}. It is thought that these differences arise due to the coupling of moist convective activity with the dynamics—i.e., these waves are more aptly described as CCERs.

Much like other large-scale tropical intraseasonal systems, such as the Madden–Julian Oscillation and the Boreal Summer Intraseasonal Oscillation, it is now clear that the evolution of moisture plays a crucial role in the QBWO. For example, the moisture mode paradigm (to describe moisture circulation feedback mechanism) has been applied to the QBWO in some regions of the Northern \citep{gonzalez2019distinct, mayta2022westward, dong2024propagation} and Southern Hemispheres \citep{yangSH, sambrita}. These regional studies were motivated by the fact that moisture anomalies and convection are almost collocated and grow in phase during the QBWO—a feature also seen in composites of CCERs \citep{nakamura2022convective}—suggesting that prognostic moisture is an essential part of the dynamics of these systems. Indeed, this line of work stems from the ideas in \cite{sobel2001weak} and has been systematically developed to include moisture coupling in models of varying complexity, with features such as evaporative feedback \citep{fuchs2019simple}, moisture gradients \citep{sukhatme2014low, suhasIVP, schrottle}, dynamically consistent quasi-geostrophic scaling \citep{joy} and background winds and shear \citep{chen2022model, mayta2022westward}. These frameworks have been applied to systems ranging from the planetary scale \citep{adames2016mjo, ahmed2021, wangsobel} to tropical synoptic lows and cyclones \citep{lahaye2016, adames2018, diaz2019, suhasboos, chaud}. In fact, the formation of moisture modes throughout the tropics, in regions of significant moisture gradients, has recently been highlighted by \cite{mayta2024stirring}.

\begin{figure}
    \centering
    \includegraphics[width=0.98\linewidth]{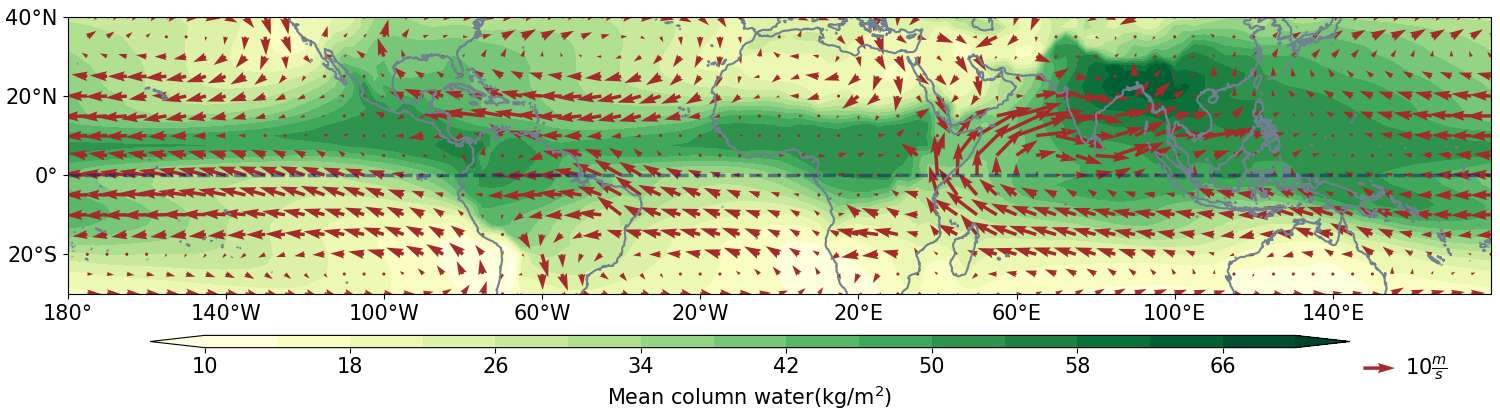}
    \caption{Mean column integrated (1000 hPa to 250 hPa) moisture and mean winds at 850 hPa during the boreal summer (JJAS).}
    \label{BG1}
\end{figure}

Despite extensive regional studies, a comprehensive understanding of how the QBWO manifests and evolves across different moisture and wind regimes—from relatively dry to moisture-dominated—remains fragmented. In this paper, we study the structure of the westward-propagating QBWO across the northern hemisphere tropics. Our area of study spans Africa (dry), the Atlantic (intermediate moisture), the West Pacific (moist) and Indo-China (very large background moisture). In particular, we make a distinction between relatively dry and moist regions based on the climatological background moisture content \citep{sukhatme2012} and background wind conditions (Figure \ref{BG1}). Notably, as seen in Figure \ref{BG1}, in the lower troposphere from the equator to the subtropics, the dry and intermediately moist regions are characterized by easterly winds, the West Pacific is a transition zone with both easterly and meridional mean flows and the very moist regions have a mean monsoonal westerly zonal flow during the boreal summer. By systematically contrasting these regimes, this work provides new insights into the coupling between circulation and moisture that sustains and modulates the QBWO across the tropics. The data used and methods employed for the analysis are described in Section 2. The basic structure of wind, moisture, and outgoing longwave radiation (OLR) anomalies across nine different tropical regions is presented in Section 3. The validity of the moisture mode idea is also checked here. From the nine regions, for brevity, we select two areas representing the dry and moist regions to illustrate the phase relation between moisture and vorticity. Vertical structures of horizontal wind, specific humidity, and pressure velocity with latitude and longitude are shown in Section 4. With a robust system description, we then contrast the vorticity and moisture budgets in the selected dry versus moist regions in Sections 5 and 6. Specifically, in section 5, we perform a vorticity budget, and along with the plots for the two chosen domains, we also elucidate the role of the dominant terms responsible for the propagation of the vorticity anomalies in the other seven regions. In Section 6, similarly, we compare our observations with a moisture budget in dry and moist regions and try to identify the key differences in the movement of moisture anomalies in these differing environments. Finally, we summarize the pivotal factors that control our results in the conclusions in Section 7.

\section{Data and methodology}

Twenty-one years (1990-2010) of daily mean of outgoing longwave radiation (OLR) from the National Oceanic and Atmospheric Administration (NOAA) (\url{https://www.ncei.noaa.gov/products/climate-data-records/outgoing-longwave-radiation-daily}) is used as a proxy for the convection. Wind and specific humidity data are obtained from ERA5 reanalysis data at different pressure levels (\url{https://cds.climate.copernicus.eu/datasets/reanalysis-era5-pressure-levels?tab=overview}) for the same duration of 1990-2010. Here, six-hourly data (four times daily) is averaged to obtain daily data. We used the column integrated data from $1000$ to $250$ hPa for the moisture budget. The vorticity is obtained from the horizontal winds using the \textit{windspharm} package \citep{dawson2016windspharm}. The vorticity budget is performed from 750 hPa to 550 hPa (middle-lower troposphere), considering the barotropic structure of vorticity and the highest value in this regime, and the average is shown. 

All the data are filtered in the 10 to 25 day time period and -1 to -10 planetary wavenumbers to represent the westward propagating QBWO. For the seasonal mean, we have applied a low-pass filter at the cut-off frequency $\frac{1}{90}$ day\textsuperscript{-1}, which means cycles of more than 90 days are considered as the seasonal mean. For filtering, we have used a basic approach using the Fourier and the inverse Fourier transforms, utilizing the \textit{FFT} and \textit{IFFT} functions of \textit{scipy}. To extract the westward QBWO, after performing a Fourier transform on the data in both time and longitude axes, the amplitudes of the relevant time frequencies and longitude frequencies (planetary wave numbers) are retained, while all other amplitudes are reduced to zero. We have applied the inverse Fourier transform to these modified amplitudes and obtained the filtered data. For extracting the seasonal mean, the Fourier and inverse Fourier transforms are applied only on the time axis.

In composites, we have chosen boxes of $10^\circ\times 10^\circ$ across the tropics, with the constant latitude band of $10^\circ\text{N}-20^\circ\text{N}$. All the boxes are centered at $15^\circ\text{N}$, and the longitude is varied across nine regions, the details of which are given below. Day 0 of the composite is chosen based on two criteria. First, on Day 0, the filtered OLR should be less than one standard deviation from the seasonal mean (All the values are box averaged, season here denotes the months of June-September, JJAS) and would be a local minimum (Day -1 and Day 1 have higher values of filtered OLR than Day 0). Secondly, the OLR should be higher than one standard deviation from the seasonal mean on any of the next 5 to 12 days from Day 0. The averaged standard deviation of the box for different regions in $\mathrm{W/m^2}$ is, East Pacific: 8.96, West Atlantic: 8.45, Central Atlantic: 7.38, West African: 7.02, Central African: 6.37, Arabian Sea: 9.30, Bay of Bengal: 10.84, West Pacific: 14.45, Central Pacific: 9.56. This captures the oscillations with a deep convection and reasonably high OLR thereafter, which is at least one half-wave. In this way, we have obtained a number of Day 0 instances in the 21 year period. Specifically, the instances vary between 37 and 58 (East pacific: 47, West Atlantic: 42, Central Atlantic: 50, West African: 58, Central African: 40, Arabian Sea: 48, Bay of Bengal: 53, West Pacific: 46, Central Pacific: 37) from one region to another, but are statistically significant for our study. Each of the variables obtained on Day 0 is then averaged, so that it can be considered as representative of that particular variable on Day 0. The same argument holds for all other lead or lag days (Day -1, Day 1, etc.).

Nine zones are selected for composites, all have latitude extent of $20^\circ\,\mathrm{N}-10^\circ\,\mathrm{N}$, longitude extents are given as: East Pacific $(130^\circ\,\mathrm{W}-120^\circ\,\mathrm{W})$, West Atlantic $(60^\circ\,\mathrm{W}-50^\circ\,\mathrm{W})$, Central Atlantic $(30^\circ\,\mathrm{W}-20^\circ\,\mathrm{W})$, West African $(0^\circ\,-10^\circ\,\mathrm{E})$, East African $(30^\circ\,\mathrm{E}-40^\circ\,\mathrm{E})$, Arabian Sea $(60^\circ\,\mathrm{E}-70^\circ\,\mathrm{E})$, Bay of Bengal $(80^\circ\,\mathrm{E}-90^\circ\,\mathrm{E})$, West Pacific $(130^\circ\,\mathrm{E}-140^\circ\,\mathrm{E})$, Central Pacific $(170^\circ\,\mathrm{E}-180^\circ)$. As mentioned, these sample fairly dry regions (such as Central Africa) and very moist regions (such as the Bay of Bengal) based on the background precipitable water \citep[Figure \ref{BG1},][]{sukhatme2012}.

\begin{figure*}[ht]
\setlength{\fboxsep}{0pt}%
\setlength{\fboxrule}{0pt}%
\begin{center}
\includegraphics[width=0.99\textwidth]{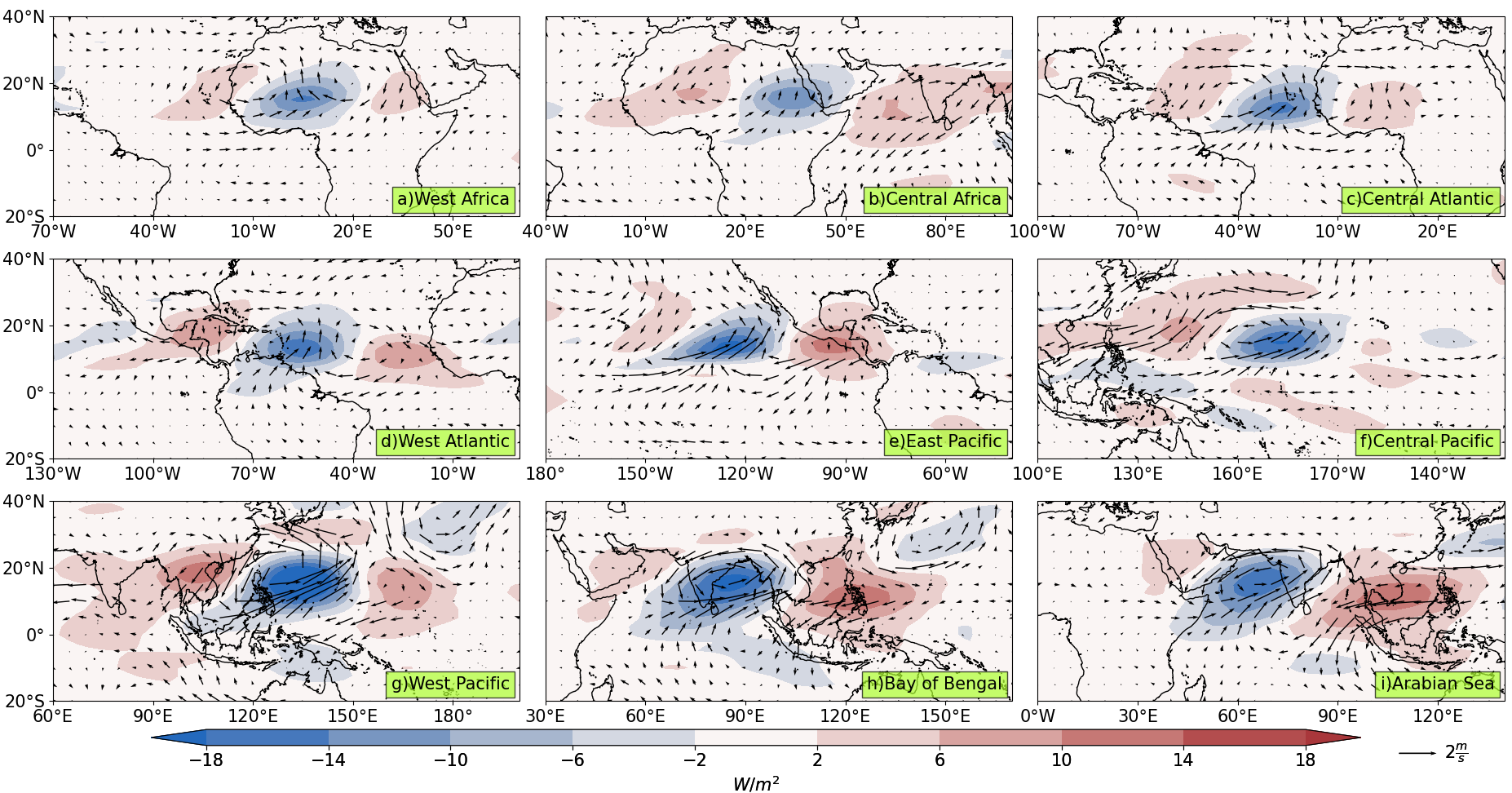}
\end{center}
\caption{Composite OLR (shading) and 850 hPa horizontal wind (quivers) anomalies on Day 0 in the northern hemisphere tropics. The mentioned regions are chosen by the methodology described before. 
}
\label{F1}
\end{figure*}

\begin{figure*}[ht]
    \centering
    \includegraphics[width=0.99\textwidth]{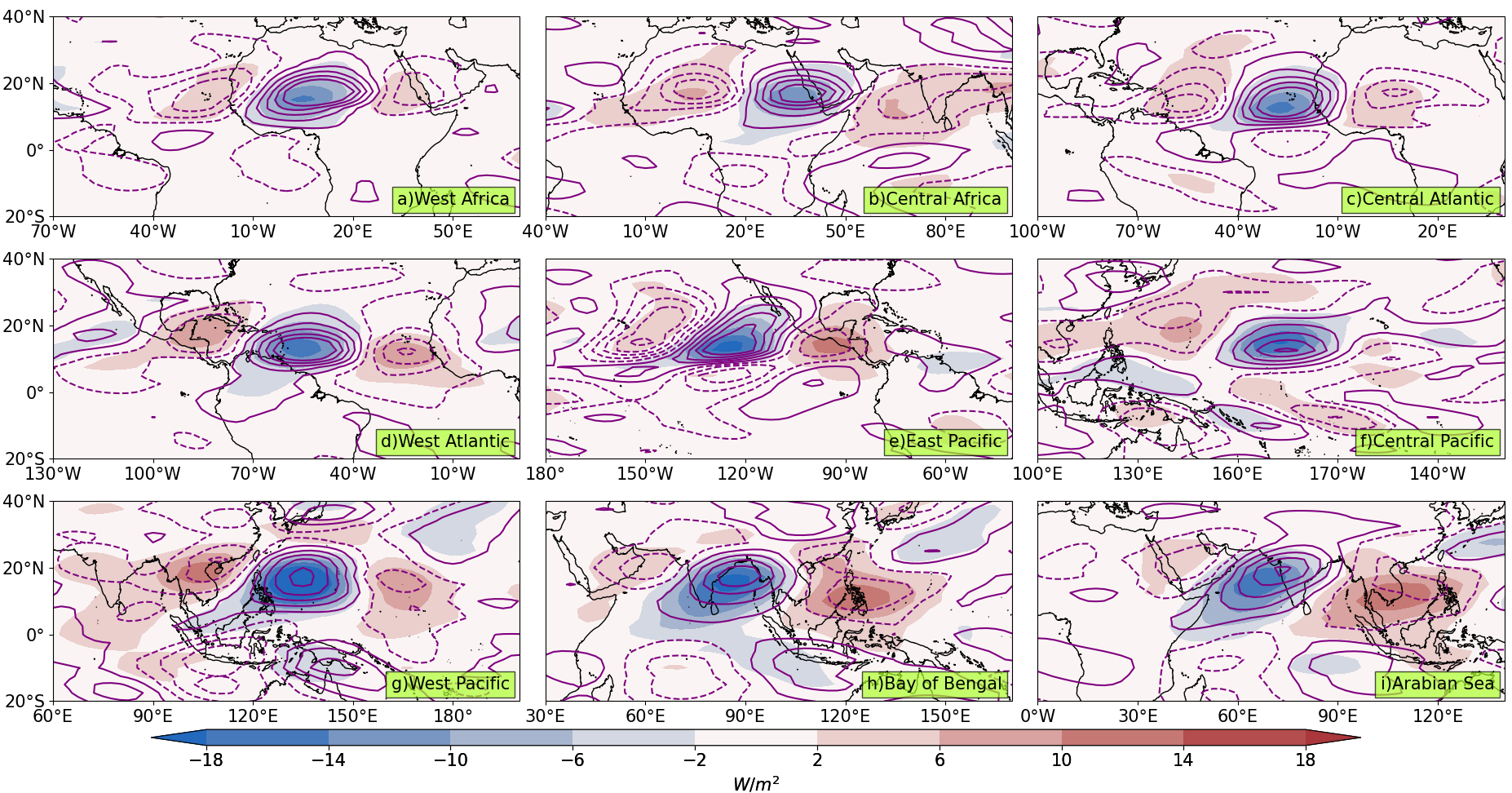}
    \caption{Composite OLR (shading) and moisture (solid and dashed contours denote positive and negative values respectively) anomalies on Day 0. Interval of contours is $\SI{0.6}{\kg/\m^2}$. The regions are the same as those in Figure \ref{F1}. 
    }
\label{F2}
\end{figure*}

\begin{figure*}[ht]
    \centering
    \includegraphics[width=0.99\linewidth]{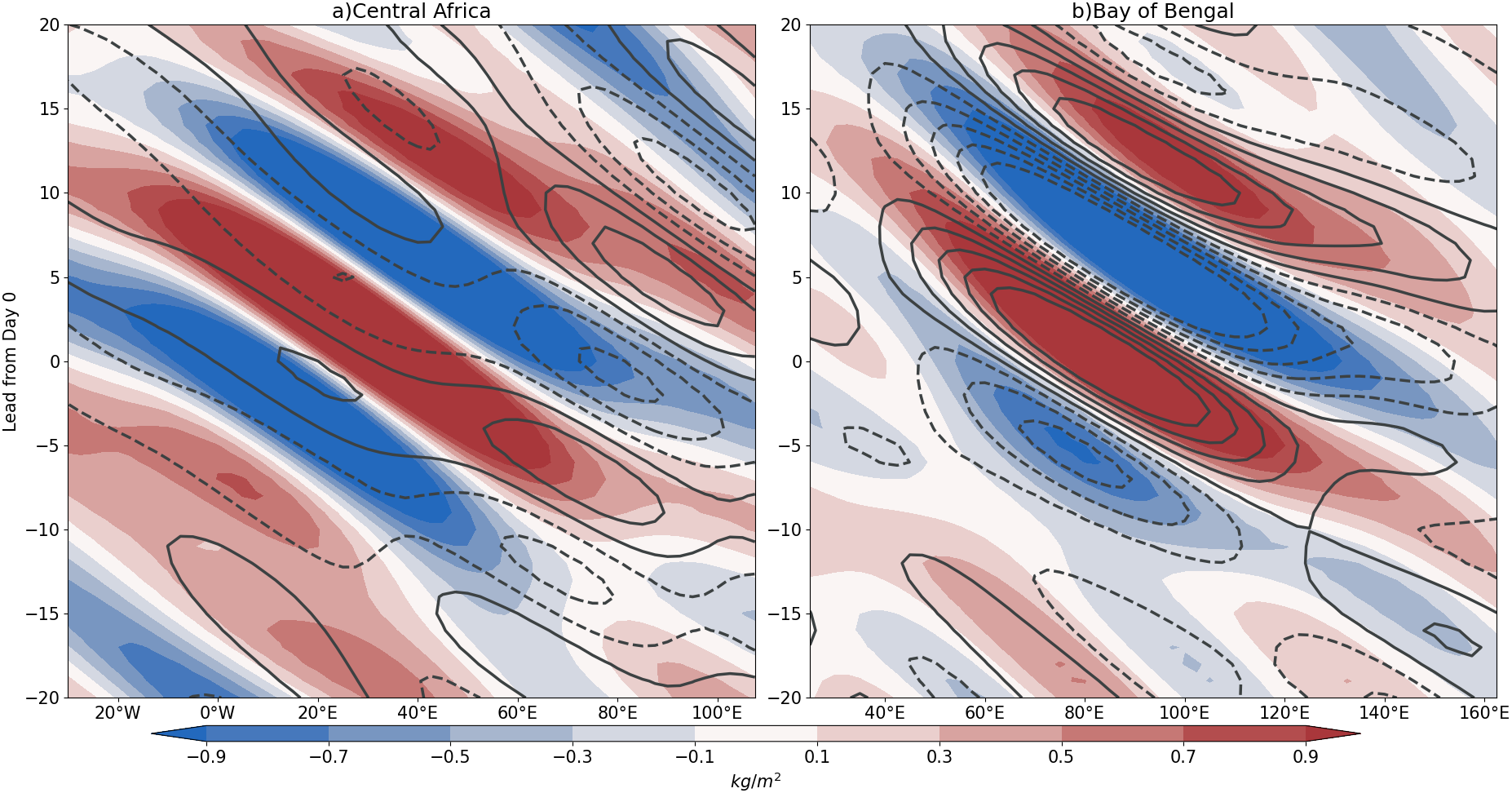}
    \caption{Hovm\"oller diagram for column moisture (color shading) and vorticity (contours, positive in solid and negative in dashed) anomalies, the difference between adjacent contours is $5\times10^{-7}\text{s}^{-1}$. Averaged values across latitudes $20^\circ\text{N}-10^\circ\text{N}$ are plotted in each case, the Y-axis shows lead/lag from Day 0. The left panel (a)) shows a composite for the dry Central African region and the right one (b)) represents a composite for the moist Bay of Bengal.}
    \label{F9}
\end{figure*}

\section{OLR, wind and column moisture anomalies}

We begin with filtered Day 0 OLR and 850 hPa wind anomalies in different regions of the tropics (Figure \ref{F1}). Note that the patterns obtained are almost identical to those shown in Figure \ref{F1} if we use a criterion of 95\% significance in a student's t-test (Figure S1). The panels are arranged such that the top most row corresponds to climatologically dry regions while the bottom row contains regions with the largest background precipitable water. We immediately notice a difference in the relative positions of the OLR anomaly and the wind gyre. Specifically, in drier regions (west Africa, Central Africa and the Central Atlantic, i.e., the top row of Figure \ref{F1}) the wind gyres and OLR anomalies are displaced with largest OLR anomalies lying on the edges of the gyres \citep[this displacement in Central Atlantic case can also be seen in][]{mayta2022westward}. Whereas, in the regions of large background moisture (west Pacific, Bay of Bengal and the Arabian Sea, i.e., the bottom row of Figure \ref{F1}) the OLR anomalies appear to be contained
within the gyres of the QBWO which is consistent with previous studies  \citep{gonzalez2019distinct,mayta2022westward,dong2024propagation}. More specifically, there is a gradual progression that can be observed as we go down the rows of Figure \ref{F1}, where the OLR anomaly is located at the edge and then gradually enters the gyre as we move from climatologically drier to more moist regions. Indeed, in the very moist regions (Bay of Bengal and the Arabian Sea), the moisture anomaly is completely contained within the QBWO gyre. A zoomed in version of Figure \ref{F1} that shows the contrasting behaviour of OLR anomalies and the gyre location in representative dry and moist regions is presented in the supplementary material (Figure S2). Moreover, the magnitude of the anomalies is larger in the moist regions and the gyres themselves are somewhat more isotropic as compared to the more longitudinally elongated structures in the drier locations. Interestingly, in all regions the OLR and moisture anomalies are collocated as seen in Figure \ref{F2}. Therefore, we can interchangeably use OLR and moisture to indicate the convection and changes in the moisture anomaly are expected to capture the change in convection. In essence, performing the moisture budget would be a reasonable manner in which to explain and understand convective activity \citep{am2018,ghatak2025northward}.

As the cyclonic gyres are regions of high vorticity, this implies the collocation of positive vorticity anomalies with positive moisture anomalies in moist regions, while in drier regions, positive vorticity anomalies are in phase with reduced column moisture anomalies. To probe the development of moisture and 850 hPa vorticity anomalies, we present a time-longitude Hovm\"oller diagram in Figure \ref{F9}. For clarity, we have shown only the Central African and Bay of Bengal regions as representative of relatively dry and moist locations, respectively, though we have checked that this pattern holds in other regions as well. Here, average values of these variables across the latitudes $20^\circ\text{N}-10^\circ\text{N}$ are shown with solid (dashed) contours indicating positive (negative) relative vorticity, while colors represent the moisture anomaly. We see the coexistence of vorticity and moisture anomalies, i.e., both have the same signs indicating that cyclonic circulation has simultaneous moisture build-up in the moist regions and a lead-lag between them in the dry regions, i.e., vorticity develops before the moisture anomaly supporting accumulation of moisture on the eastward side of the cyclonic gyre. Moreover, probing the relation between vorticity and moisture anomalies a little more, we find that a scatter plot (not shown) between the two variables yields a clear linear relationship over the Bay of Bengal (indeed, this is true over the Arabian Sea too which is the other very moist region in our analysis). Thus, vorticity and moisture grow together on the timescale of the QBWO and are tightly coupled in their character in the very moist regions of the northern hemisphere tropics.

\begin{figure}[ht]
    \centering
    \includegraphics[width=0.99\linewidth]{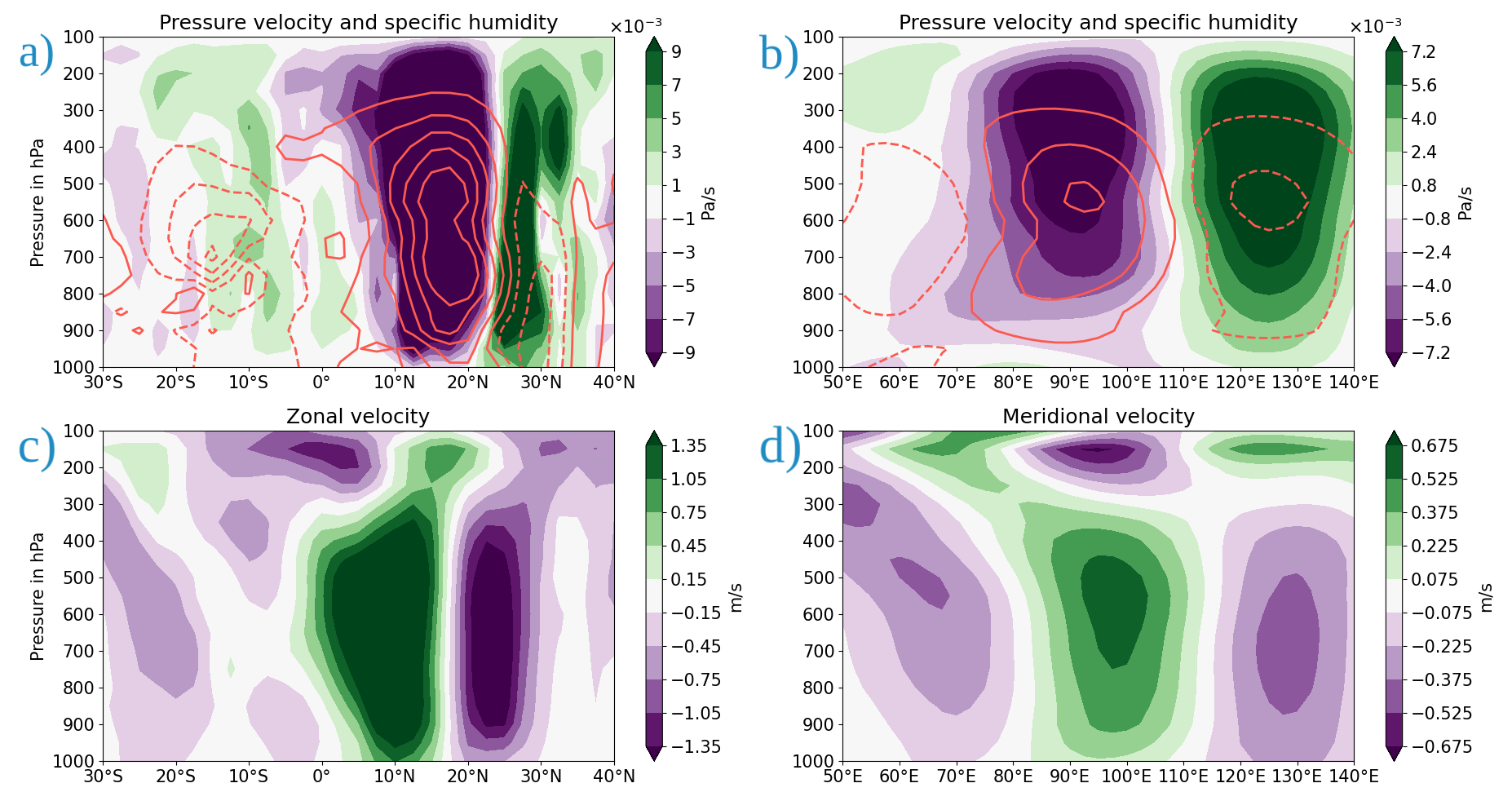}
    \caption{Vertical structure of pressure velocity (shading, upper row) and moisture (contours, upper row) with latitude (a)) and longitude (b)) on Day 0 of the QBWO composite over the Bay of Bengal. Solid (dashed) contours denote positive (negative) moisture anomalies, and the contour interval is \num{8d-5}\unit{kg/kg}. The lower panels show the vertical structures of zonal velocity with latitude (c)) and meridional velocity with longitude (d)) in the first and second column, respectively.} 
    \label{F3}
\end{figure}

\begin{figure}[ht]
    \centering
    \includegraphics[width=0.99\linewidth]{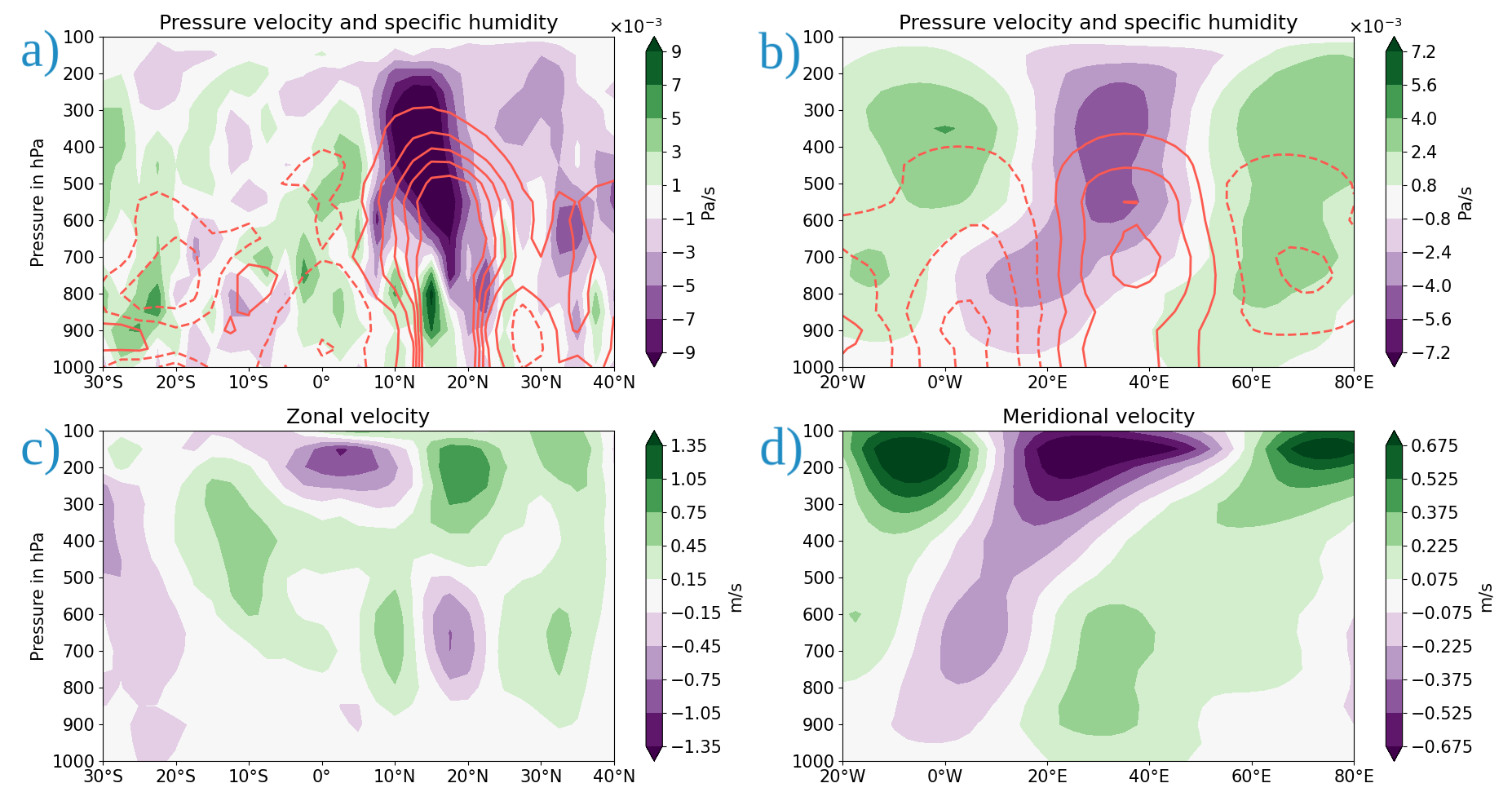}
    \caption{Same as the Figure~\ref{F3}, but the composite is over the Central African region.} 
    \label{F4}
\end{figure}

We now examine the vertical structure of the westward propagating QBWO. Here too, we use the Bay of Bengal and the Central African regions for illustration as they are largely representative of the moist and dry regions, respectively. In the moist Bay of Bengal region, for the zonal dependence, we averaged the structures over the latitudes $5^\circ \text{N}-35^\circ \text{N}$ and for the meridional profile, we have averaged the structures over the longitudes $50^\circ\text{E}-140^\circ\text{E}$. Here, specific humidity and vertical (pressure) velocity exhibit an upright or equivalent barotropic structure (Figure \ref{F3}a,b). The largest specific humidity anomaly is found between $800\text{ hPa}-500\text{ hPa}$ and while the pressure velocity is top-heavy and attains a maximum around $300\text{ hPa}$, which is broadly in accord with the composite structures of CCERs \citep{nakamura2022convective}. The zonal and meridional velocity also maintains its sign up-to the middle troposphere (Figure \ref{F3}c,d) which is consistent with observations of QBWOs from the West Pacific as noted by \citep{gonzalez2019distinct}. In addition, there are signs of a flow reversal above $300\text{ hPa}$, which is particularly clear in the meridional velocity, suggesting a first baroclinic structure as was noted for the QBWO in the West Pacific \citep{chen2010characteristics}. These features also largely agree with CCER composites at 500 hPa and 200 hPa in \cite{nakamura2022convective}, though they have not explicitly shown the sign reversal at 300 hPa. Similarly, earlier work \citep{kw1995} also noted a nearly equivalent barotropic structure of the horizontal wind gyres in CCER waves.

\section{Vertical Structure}

The composite from dry regions (Figure~\ref{F4}; we have kept the colourbars the same as in Figure \ref{F3} to facilitate comparisons between the moist and dry regions) is largely consistent when compared to the moist region, though the signature of anomalies is weaker and there are some differences as noted below. Here, for the zonal dependence, we have averaged the structures over the latitudes $5^\circ \text{N}-35^\circ \text{N}$ and for the meridional profiles, we have averaged over the longitudes $20^\circ\text{W}-80^\circ\text{E}$. The specific humidity has a columnar structure (Figure \ref{F4}a,b), and the largest anomalies are observed near 700 hPa. The pressure velocity has an eastward tilt in the lower troposphere and is upright above that as seen in the longitude-height plot (Figure \ref{F4}b). 
The zonal and meridional velocity anomalies also become clear above 700-800 hPa and show a baroclinic structure with a sign change at about 300 hPa (Figure \ref{F4}c,d) as for the moist composite. We interpret this as a baroclinic structure rather than a tilted vortex due to the upright nature of the vertical velocity in the mid to upper troposphere. Notably, in comparison to the moist regions, the zonal flow is quite confined while the meridional flow is much broader in nature. Further, the meridional velocity is considerably larger in magnitude in the upper troposphere. In essence, for the dry region, the signature of the velocity field anomalies really appears to emerge clearly above about 800 hPa which is contrast to moist composite where most of the structure extended to the surface. But, in both regions, a first baroclinic mode structure along with upright vertical wind and moisture anomalies appear to be characteristic of the QBWO. Moreover, moist static energy anomalies (not shown) are dominated by the contribution from moisture in both the relatively dry and moist regions.

\begin{figure*}[ht]
\centering
  \includegraphics[width=0.98\textwidth]{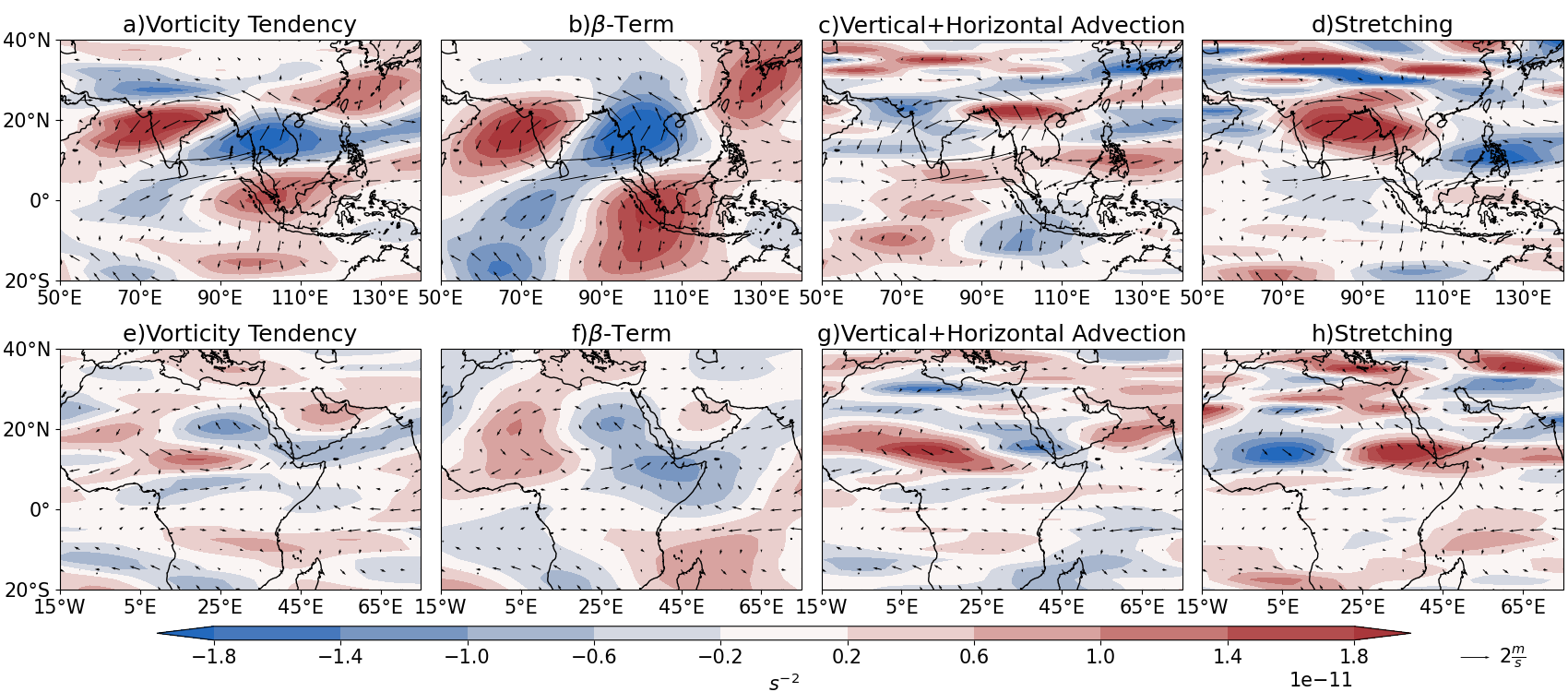}
  \caption{Important terms in the vorticity budget, averaged between 750 hPa and 550 hPa on Day 0 for the moist Bay of Bengal (upper row, a), b), c), d)) and dry Central African (lower row, e), f), g), h)) regions. The arrows represent the QBWO horizontal wind anomalies at 650 hPa.}
  \label{F6}
\end{figure*}

\section{Vorticity budget}
To examine the dynamical processes governing the QBWO, we first analyse the vorticity budget to identify the terms responsible for driving and maintaining the circulation anomalies across different moisture regimes. Denoting the vertical component of the relative vorticity by $(\zeta\equiv \frac{\partial v}{\partial x}-\frac{\partial u}{\partial y} )$, its evolution in pressure coordinates can be expressed as the horizontal divergence of a flux \citep{Haynes}, but it is usually expressed in terms of advection, stretching and tilting terms as \citep{matthews2025vorticity},
\begin{align}
\label{vort_a}
    \underbrace{\frac{\partial \zeta}{\partial t}}_\text{Tendency}=\underbrace{{-\left(u\frac{\partial }{\partial x}+v\frac{\partial }{\partial y}+\omega \frac{\partial}{\partial p}\right)\zeta}}_\text{Advection} -\underbrace{{\left(\zeta+f\right)\left(\frac{\partial u}{\partial x}+\frac{\partial v}{\partial y}\right)}}_\text{Stretching}-\underbrace{{\left(\frac{\partial \omega}{\partial x}\frac{\partial v}{\partial p}-\frac{\partial \omega}{\partial y}\frac{\partial u}{\partial p}\right)}}_\text{Tilting}-\underbrace{{\beta v}}_\text{Beta}.
\end{align}
Here $u,v,\omega$ are components of the velocity field, $f$ is the Coriolis parameter and $\beta=\frac{df}{dy}$ is its meridional gradient. In Equation \ref{vort_a}, we have separated out the $\beta$ term which also comes from horizontal advection. Now, in budget calculations for a particular range of spatio-temporal scales (in our case, the QBWO), 
the form of the vorticity equation used is \citep{ghatak2025northward},
\begin{align}
\label{vort}
    \left\langle\frac{\partial \zeta}{\partial t}^\prime\right\rangle={\left\langle-\left(u\frac{\partial }{\partial x}+v\frac{\partial }{\partial y}+\omega \frac{\partial}{\partial p}\right)\zeta\right\rangle}^\prime -{\left\langle\left(\zeta+f\right)\left(\frac{\partial u}{\partial x}+\frac{\partial v}{\partial y}\right)\right\rangle}^\prime-{\left\langle\left(\frac{\partial \omega}{\partial x}\frac{\partial v}{\partial p}-\frac{\partial \omega}{\partial y}\frac{\partial u}{\partial p}\right)\right\rangle}^\prime-{\langle\beta v\rangle}^\prime+ \\ \textrm{Residue} \nonumber.
\end{align}
Here, primed quantities are the QBWO filtered anomalies. Equation \ref{vort} has a residue which contains neglected scale interactions that can potentially influence the QBWO vorticity. In all nine regions, we compute the vorticity budget in the middle-lower troposphere (750 hPa to 550 hPa, with 50 hPa intervals, total 5 layers). Specifically, we have taken the average (denoted by angular brackets) across the five layers to identify the factors that influence the tendency of the QBWO vorticity. Though the vorticity is predominantly barotropic, its maxima are located at around 600 or 650 hPa in both dry and moist regions, due to which we have taken the average value of the above-mentioned  pressure levels.

The vorticity budget is shown in Figure \ref{F6} --- here too, the Bay of Bengal and Central Africa are representative of the budgets in moist and dry regions, respectively. The tendency immediately indicates a westward propagation of the QBWO gyre. As expected, in the moist (Bay of Bengal) and dry (Central Africa) regions, the $\beta$-term matches qualitatively with the vorticity tendency. But, in both cases, the $\beta$-term is of a much larger scale and has a different tilt than the tendency itself, suggesting that it is tempered by some of the other terms in Equation \ref{vort}. The stretching term has the largest magnitude but does not match the tendency --- in fact, it is aligned with horizontal convergence which in the moist regions is contained within the QBWO gyre and is on the edges of the gyre in the dry regions. Thus, in relatively dry regions, stretching largely counters the $\beta$-effect, which is in accord with theoretical expectations for an ER considering convergence towards the westward side of the gyre and the divergence towards the eastward side \citep{matthews2025vorticity}. Whereas, in moist regions, 
as pointed out in the context of CCERs \citep{matthews2025vorticity}, its role seems more to help in the growth of the vorticity anomaly on Day 0 rather than in the propagation of the QBWO gyre. 

The advection of vorticity (especially horizontal advection, not shown separately but is the dominant part of this term) appears to oppose stretching --- this is especially clear in the dry composite and also to an extent in the moist region. Such a cancellation of terms in the vorticity budget has been discussed by \cite{SH1985} on seasonal timescales and has also been noted in other phenomena such as tropical lows \citep{boos_drift,pradeep2}; indeed, \cite{matthews2025vorticity} found horizontal advection to lead to the decay of the vorticity anomaly on Day 0 in CCERs. Thus, in the budget of dry regions, advection supports the tendency. While in the moist case, advection aids in the decay of the existing vorticity and also appears to oppose the vorticity tendency towards the edges of the gyre.

Noting that the tilting term in Equation \ref{vort} is negligible throughout, indicating small horizontal vorticity components and a lack of a systematic gradient in the vertical velocity, and that the vertical advection of relative vorticity is also very small, a simplified form for Equation \ref{vort} for further analysis reads,
\begin{align}
\label{vort1}
    \left\langle\frac{\partial \zeta}{\partial t}^\prime\right\rangle={\left\langle-\left(u\frac{\partial }{\partial x}+v\frac{\partial }{\partial y}\right)\zeta\right\rangle}^\prime -{\left\langle\left(\zeta+f\right)\left(\frac{\partial u}{\partial x}+\frac{\partial v}{\partial y}\right)\right\rangle}^\prime-{\langle\beta v\rangle}^\prime+  \textrm{Residue}.
\end{align}

\begin{figure}[hp]
    \centering
    \includegraphics[width=0.98\linewidth]{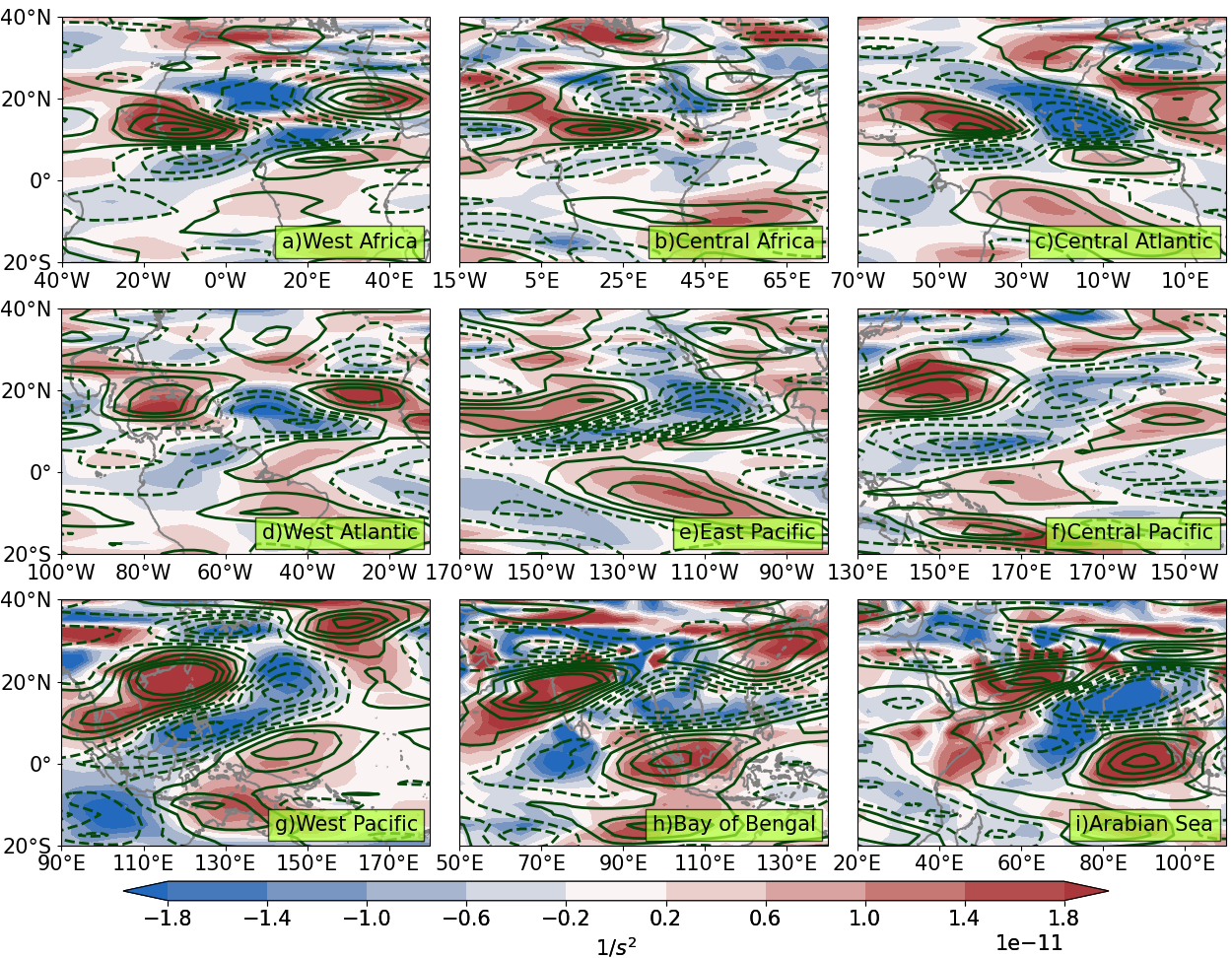}
    \caption{Sum of dominant terms (as mentioned in the Table \ref{tab:T1}) in vorticity budget across the nine regions in the tropics on Day 0. Solid (dashed) contours indicate  positive (negative) vorticity tendency with contour intervals of \num{4d-12}\unit{s^{-2}}.}
    \label{F7}
\end{figure}
In the above, the residue now includes the missing scale interactions and also the neglected tilting and vertical advection terms. A map of the residue for the typical dry and moist regions in shown in Figure S6 --- as is noted, this is quite small and unorganized in character in the regions where the tendency is most significant. This suggests that the other terms in the RHS of Equation \ref{vort1} are a reasonable approximation to the vorticity anomaly tendency.

To understand the role of horizontal advection and stretching in more detail, we now split the terms in the vorticity budget into the effects of mean and filtered parts of the fields, in particular, our intent is to examine deviations from a Sverdrup-like balance \citep{gill1980some}, i.e., given that $\zeta' \!\ll \!f$ (which is a reasonable approximation in most of the regions in our study), how well does the planetary component of stretching $\left(-f(\frac{\partial u}{\partial x}+\frac{\partial v}{\partial y})=f\frac{\partial \omega}{\partial p}\right)$, advection by the mean flow along with the $\beta$ term approximate the tendency. Thus, horizontal advection is approximately decomposed into mean and perturbed fields as follows. The perturbed field (denoted by a prime) is, as before, the westward quasi-biweekly oscillation, and the mean (denoted by an overbar) is defined as oscillations with time period of 90 days or more, i.e., it roughly captures the seasonal average. Neglecting the residue and the product of two primed fields, this yields,
\begin{align}
    \left(\left(u\frac{\partial }{\partial x}+v\frac{\partial }{\partial y}\right)\zeta\right)^\prime=\bar{u}\frac{\partial \zeta^\prime}{\partial x}+u^\prime\frac{\partial \bar{\zeta}}{\partial x}+\bar{v}\frac{\partial \zeta^\prime}{\partial y}+v^\prime\frac{\partial \bar{\zeta}}{\partial y}.\label{vort3}
\end{align}

In the dry and intermediately moist regions, i.e., Central Africa, West Africa, Central Atlantic, West Atlantic, East Pacific and Central Pacific regions (those in the first two rows of Figure \ref{F1}), planetary stretching and mean flow advection of the vorticity anomaly are dominant in their respective terms, and as shown in the first two rows of Figure \ref{F7}, $f \nabla\!\cdot\!{\bm u}', \beta v'$ and $\bar{u}\zeta'_x + \bar{v} \zeta'_y$ together give a reasonable estimate of the tendency. This is akin to the balance obtained for flanking Rossby waves in a shallow water experiment that examined the stationary response to steady forcing in the presence of a zonal jet \citep{mont}. While $\beta$ (stretching) consistently aids (hinders) the movement of the vorticity anomaly, there are differences in the contributions of two advective terms. For example, over the Eastern Pacific, West Africa, Central Atlantic and West Atlantic, only zonal advection plays a significant role in supporting the tendency, while in the other two locations (Central Africa and Central Pacific) meridional advection is also important. In effect, as per the details in Table \ref{tab:T1}, $\beta$, planetary stretching and mean flow advection of the QBWO vorticity anomaly approximate the Day 0 vorticity tendency in dry and intermediately moist regions of the northern hemisphere tropics. The error in these simple approximations (as compared to the complete RHS of Equation \ref{vort1}) is also listed in Table \ref{tab:T1} and is consistently less than about 20\% in all the relatively dry regions.

As for the three moist regions (last row of Figure \ref{F7}; third column of Table \ref{tab:T1}), in the West Pacific, the aforementioned combination of planetary stretching, horizontal advection by the mean flow and $\beta$ works well in describing the tendency but it is the meridional advection of the QBWO vorticity anomaly ($\bar{v} \zeta'_y$) that is dominant. In the Arabian Sea and Bay of Bengal regions, the situation is markedly different, wherein the advection of the mean vorticity by the eddy winds with planetary stretching term shape $\beta$ so as to approximate the tendency. While stretching is contained in the QBWO gyre and is associated with growth on Day 0, we find that it is required with advection and $\beta$ when considering the propagation of the vorticity anomaly. Therefore, as we move from dry to very moist regions, the most important aspect of horizontal advection changes from mean flows advecting vorticity anomalies to the anomalous velocity advecting the mean vorticity.

\begin{table}
\begin{center}
\caption{Dominant terms in the vorticity tendency of all the nine regions in addition to $-\beta v'$. Note that ${\bm u}'=(u',v'),  \bar{{\bm u}}=(\bar{u},\bar{v})$.}
    \label{tab:T1}
\begin{scriptsize}
\begin{tabular}{ |c|c||c|c||c|c| } 
 \hline
 Region (\% error)& Dominant terms & Region (\% error) & Dominant terms & Region (\% error) & Dominant terms\\
 \hline
 W. Africa (18.56\%)& $-\bar{u}\frac{\partial \zeta^\prime}{\partial x} + f\nabla\!\cdot\!{\bm u}'$ &W. Atlantic (21.38\%)& $-\bar{u}\frac{\partial \zeta^\prime}{\partial x} + f\nabla\!\cdot\!{\bm u}'$ &Arabian Sea (29.09\%) &$-{\bm u'}\!\cdot \!\nabla \!\bar{\zeta}+f \nabla\!\cdot\!{\bm u}'$ \\
 C. Africa (14.85\%)& $-\bar{{\bm u}}\!\cdot \!\nabla \zeta'+ f\nabla\!\cdot\!{\bm u}'$&C. Pacific (12.81\%)& $-\bar{{\bm u}}\!\cdot \!\nabla \zeta'+f\nabla\!\cdot\!{\bm u}'$ & Bay of Bengal (19.17\%) &$-{\bm u'}\!\cdot \!\nabla \!\bar{\zeta}+f \nabla\!\cdot\!{\bm u}'$\\
 C. Atlantic (20.88\%)& $-\bar{u}\frac{\partial \zeta^\prime}{\partial x}+f\nabla\!\cdot\!{\bm u}'$&E. Pacific (16.22\%)& $-\bar{u}\frac{\partial \zeta^\prime}{\partial x}+f\nabla\!\cdot\!{\bm u}'$ &W. Pacific (12.36\%)&$-\bar{v}\frac{\partial \zeta^\prime}{\partial y}+f\nabla\!\cdot\!{\bm u}'$\\
 \hline
\end{tabular}
\end{scriptsize}
\end{center}
\end{table}

\begin{figure*}[hp]
    \centering
    \includegraphics[width=0.98\textwidth]{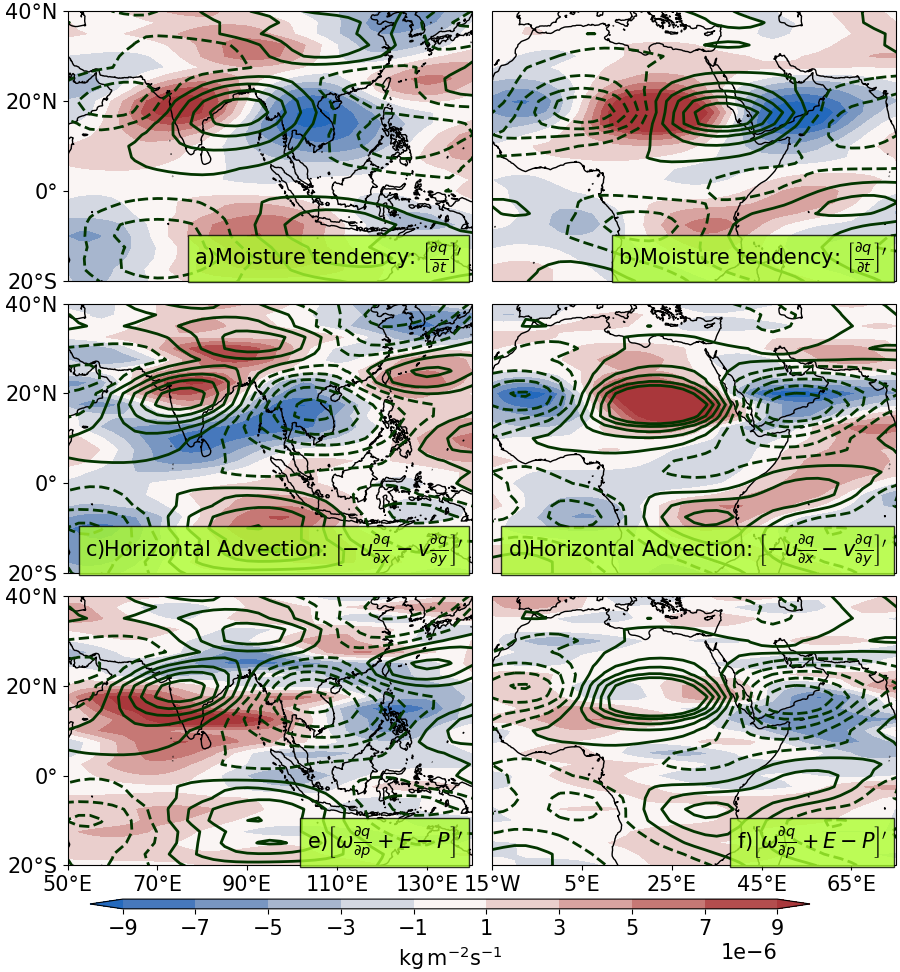}
    \caption{Terms in the moisture budget on Day 0 for the Bay of Bengal (left) and Central Africa (right). The first row (a),b)) shows the moisture tendency (shading) and column integrated moisture anomaly (contours with intervals of \num{0.6} \unit{\mathrm{kg\,m^{-2}}}). Horizontal advection and the sum of vertical advection, evaporation and precipitation are shown in colors in the second (c),d)) and third (e),f)) rows, respectively, along with contours of moisture tendency (The contour intervals are same as the colorbar, \num{2d-6} \unit{\mathrm{kg\,m^{-2}\,s^{-1}}}). In all plots, solid contours are for positive and dashed for negative values of a variable.}
    \label{F8}
\end{figure*}

\section{Moisture budget}
To complement the dynamical analysis, we examine the moisture budget, which reveals that the processes controlling the distribution and evolution of moisture anomalies associated with the QBWO differ across dry and moist regions. The total moisture in a vertical column of the atmosphere evolves as follows \citep{am2018,ghatak2025northward}, where primed quantities denote anomalous fields as in the vorticity budget.
\begin{align}
\underbrace{\frac{\partial [q^\prime]}{\partial t}}_\text{Tendency Term}=\underbrace{-\left[u\frac{\partial q }{\partial x}\right]^\prime-\left[v\frac{\partial q}{\partial y}\right]^\prime}_\text{Horizontal advection}-\underbrace{\left[\omega \frac{\partial q}{\partial p}\right]^\prime}_\text{Vertical advection} -P^\prime+E^\prime+ \textrm{Residue}.
\label{E2}
\end{align}
Here, precipitation ($P$) and evaporation ($E$) act as a sink and source of moisture, respectively. Square brackets indicate integration across the pressure levels. Specifically, we have performed the moisture budget (integrated specific humidity from 1000 hPa to 250 hPa with a 50 hPa step size) on Day 0 to investigate the possible causes for the evolution of the QBWO moisture anomaly. 

In the nine regions considered, the moist and dry locations behave in a largely consistent manner, and we represent them by the Bay of Bengal and Central Africa, respectively. As with vorticity, the moisture anomaly tendency immediately suggests a coherent westward movement of the moisture anomaly (first row of Figure \ref{F8}). Moreover, as seen in the second row of Figure \ref{F8}, in both moist and dry regions, horizontal advection contributes significantly to the moisture tendency term. Indeed, in the dry region composite (Figure \ref{F8}; second column), horizontal advection (panel d)) almost completely explains the tendency. Whereas, in moist regions, vertical advection $-P +E$ also plays a role. For example, in the Bay of Bengal (Figure \ref{F8}; panel e)), the difference of vertical moisture advection, precipitation and evaporation helps contribute to the tendency on the southwestern edge of the existing gyre. The role of this term in the Central African composite is quite small, and it only appears to aid in the decay of the existing anomaly. 

\begin{figure*}[hb]
\centering
  \includegraphics[width=0.98\textwidth]{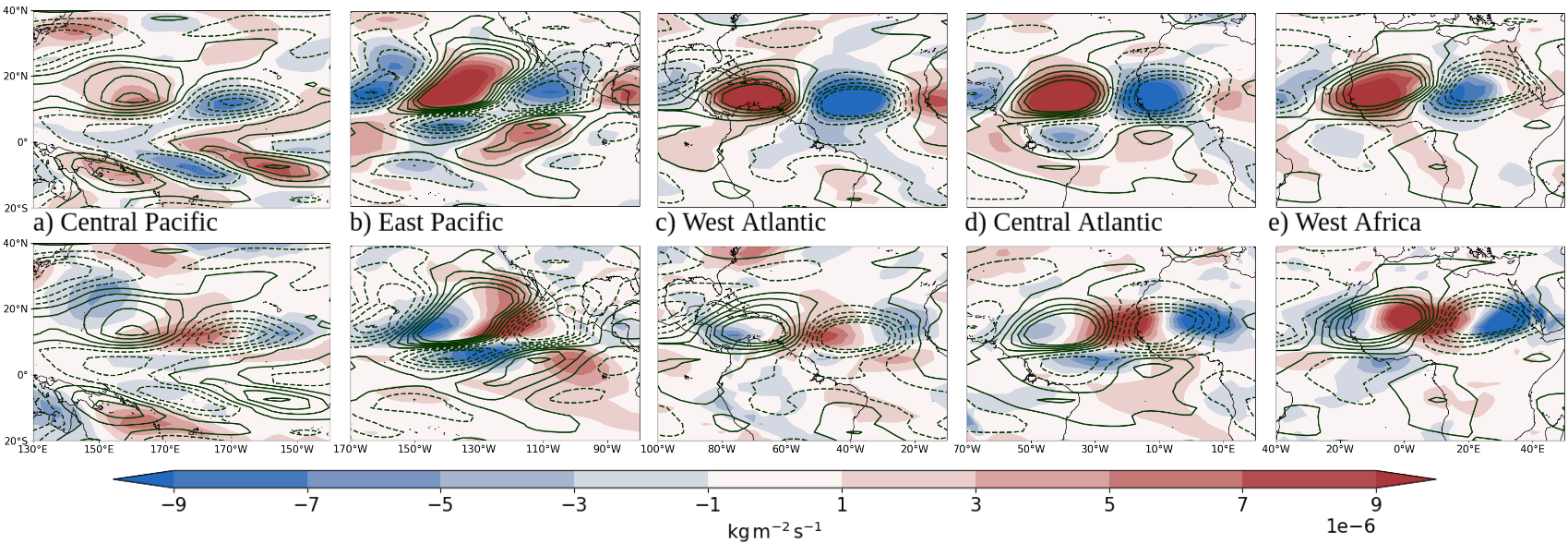}
  
  \caption{Advection of the moisture anomaly by the zonal mean wind ( $\left[\bar{u}\frac{\partial q^\prime}{\partial x}\right]$, first row) and advection of the background moisture by the eddy meridional wind ( $\left[v^\prime\frac{\partial \bar{q}}{\partial y}\right]$, second row) with moisture tendency in contours. The columns correspond to the central Pacific, east Pacific, west Atlantic, central Atlantic and west African regions (from left to right), respectively. \label{S3}} 
\end{figure*}

\begin{figure*}[hp]
    \centering
    \includegraphics[width=0.98\textwidth]{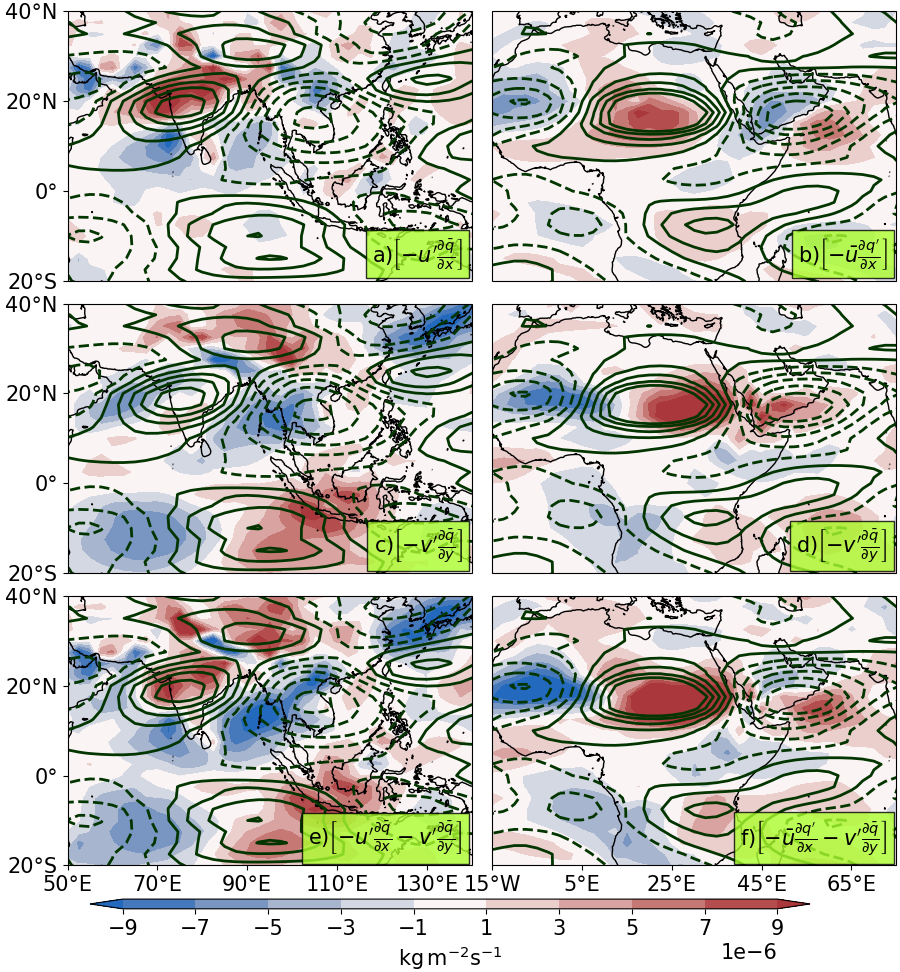}
    \caption{Decomposition of the horizontal advection of the moisture into combinations of mean and perturbed variables on Day 0 in moist Bay of Bengal (left column) and dry Central African (right column) regions. Left panels show advection of mean moisture by perturbed zonal (a)), perturbed meridional (c)) winds and addition of the two (e)). Right panels show advection of perturbed moisture by zonal mean wind (b)) and mean moisture by perturbed meridional wind (d)) and addition of the two (f)). Positive (negative) moisture tendency is overlain using solid (dashed) contours in all panels. Only the terms that chiefly control the moisture tendency are plotted.}
    \label{F10}
\end{figure*}

Given its importance, to better the horizontal advection in the moisture budget, we decompose it into products of mean and filtered variables. In the Central African region (second column of Figure~\ref{F10}), the largest terms are the zonal advection of the moisture anomaly by the mean wind $([\Bar{u}\frac{\partial q'}{\partial x}])$ and the meridional advection of the mean moisture by the QBWO wind anomaly $([v'\frac{\partial \Bar{q}}{\partial y}])$. Physically, the meridional wind anomaly acts on the large equator to pole moisture gradient alongside which the weak background zonal wind advects the moisture anomaly which has a significant longitudinal gradient --- this can be seen from Figure S3 where we show the background moisture and eddy velocity (Figure S3a) and the moisture anomaly with the seasonal background flow (Figure S3b) in the Central African region. As shown in Figure \ref{S3}, this is also true for the West African, Central Atlantic, West Atlantic, East Pacific and Central Pacific regions, i.e., all the relatively dry and intermediately moist locations (i.e., those shown in the first two rows of Figure \ref{F1}). In fact, in the West and Central Atlantic and East and Central Pacific regions, zonal advection itself is sufficient to explain much of the tendency \citep[indeed, the dominance of zonal mean flow advection of intraseasonal anomalies has been noted in the context of the moist static energy in the Central Atlantic region,][]{mayta2022westward}. The action of these two advecting terms in all of these regions can be understood by maps similar to Figure S3 for the region in consideration. 

In the moist Bay of Bengal region (first column of Figure \ref{F10}), the prominent terms are the zonal and meridional advection of the mean moisture by the QBWO wind anomalies, i.e., $([u'\frac{\partial \Bar{q}}{\partial x}])$ and $([v'\frac{\partial \Bar{q}}{\partial y}])$. The importance of the meridional advection of the background moisture was noted by \cite{chen2024}, but here, similar to the situation in the southwest Indian Ocean \citep{yangSH,sambrita}, we find zonal advection to also play an important role. Indeed, this is a region where both the pole to equator and east to west gradients of background moisture are present \citep{AWM,wangsobel,ghatak2025northward}. Consequently, the zonal and meridional advection of the background moisture by the QBWO wind anomalies are significant contributors to the tendency --- this can be seen in Figure S4, which shows the QBWO horizontal wind anomalies on Day 0 along with the background moisture in this region. 
In the West Pacific, the importance of horizontal advection has been noted \citep{gonzalez2019distinct}, and moisture budgets indicate mean winds advecting anomalous moisture to prominent \citep[see Figure S5,][]{dong2024propagation}. Thus, in the moist regions (Arabian Sea, Bay of Bengal and the West Pacific), different components of the horizontal advection become dominant (horizontal advection for the Arabian Sea is also shown in Figure S5). However, as noted, the difference between vertical advection of moisture (mainly comprised of $\left[\omega^\prime\frac{\partial \bar{q}}{\partial p}\right]$), precipitation, and evaporation is required to completely explain the moisture tendency in these regions. Specifically, as shown in the second row of Figure S5, in the Arabian Sea and West Pacific regions, as with the Bay of Bengal, this term aids the moisture tendency in the southwest sector of the original gyre. In all, the dominant terms contributing to horizontal moisture advection in every region are listed in Table \ref{tab:T2}. Here too, as with the vorticity budget, we observe that in the dry and intermediately moist regions, advection of moisture anomalies by the mean flow is important, while in the very moist zones, the eddy advection of background moisture is the dominant contributor to horizontal advection.

\begin{table}
\begin{center}
\caption{Dominant terms in horizontal moisture advection for all the nine regions.}
    \label{tab:T2}
\begin{small}
\begin{tabular}{ |c|c||c|c||c|c| } 
 \hline
 Region & Dominant terms & Region & Dominant terms & Region & Dominant terms\\
 \hline
 West Africa& $-\bar{u}\frac{\partial q'}{\partial x}-v'\frac{\partial \bar{q}}{\partial y}$ &West Atlantic& $-\bar{u}\frac{\partial q'}{\partial x}$ &Arabian Sea &$-u^\prime\frac{\partial \bar{q}}{\partial x}-v^\prime\frac{\partial \bar{q}}{\partial y}$ \\
 Central Africa& $-\bar{u}\frac{\partial q'}{\partial x}-v^\prime\frac{\partial \bar{q}}{\partial y}$&Central Pacific& $-\bar{u}\frac{\partial q'}{\partial x}$ & Bay of Bengal &$-u^\prime\frac{\partial \bar{q}}{\partial x}-v^\prime\frac{\partial q}{\partial y}$\\
 Central Atlantic& $-\bar{u}\frac{\partial q'}{\partial x}$&East Pacific& $-\bar{u}\frac{\partial q'}{\partial x}$ &West Pacific &$-\bar{u}\frac{\partial q'}{\partial x}-\bar{v}\frac{\partial q'}{\partial y}$\\
 \hline
\end{tabular}
\end{small}
\end{center}
\end{table}

\section{Conclusions and Discussion}

We have studied the westward propagating Quasi-Biweekly Oscillation (QBWO) during the boreal summer (JJAS). In particular, based on the background precipitable water and lower level mean winds (Figure \ref{BG1}), we examine the horizontal and vertical structure along with its vorticity and moisture budgets, across different moisture regimes in the northern hemisphere tropics: relatively dry regions over Africa, intermediate-moisture regions over the Atlantic, the wetter West Pacific, and very moist regions over the Bay of Bengal and Arabian Sea.

An immediate difference between the dry and moist regions is the position of vorticity and outgoing longwave radiation (OLR) anomalies. In the former, these anomalies are in quadrature, but as one moves to the moist regions, the OLR anomalies are observed to lie within the QBWO gyre, i.e., they are in phase with the vorticity anomaly. Moreover, there is a gradual shift in the position of the OLR and vorticity anomalies, i.e., they go from being in quadrature to in phase as the background moisture increases. Notably, moisture and OLR anomalies are always collocated irrespective of the background moisture. In fact, examining the growth of vorticity and moisture anomalies, we note that the two grow together and are linearly correlated in the moist region, whereas the vorticity builds up before the moisture anomaly in dry composites. The vertical structure of the QBWO is largely similar in all the regions, albeit with a larger magnitude of moisture and velocity anomalies in the moist regions. Specifically, moisture and vertical velocity are upright with maxima in the middle and upper troposphere, respectively. The winds suggest a first baroclinic mode form with a reversal of signs at about 300 mbar. These features, especially in the moist regions, are in accord with previous observations of the QBWO and also convectively coupled equatorial Rossby waves.

We then explored the propagation of the QBWO using vorticity and moisture budgets. The moisture-vortex structure moves westward, maintaining the lead-lag relation in the respective dry and moist regions. However, the propagation mechanisms in both regions show stark differences. In both dry and moist regions, the $\beta$-term helps in westward movement due to the configuration of the vortex. But, the $\beta$ term is quite broad and widespread as compared to the more compact nature of the vorticity tendency, i.e., by itself this term would lead to a spreading out of the QBWO vorticity anomaly. Horizontal advection aids the westward movement, while the stretching (mainly planetary) opposes the movement in the dry regions. Moreover, a decomposition reveals that the strong mean easterly winds in most dry regions dominate the horizontal advection of vorticity. Thus, in dry and intermediately moist regions, a combination of planetary stretching, mean flow advection of the vorticity anomaly, and the $\beta$-term explains much of the vorticity tendency. In the very moist regions (Arabian Sea and Bay of Bengal), planetary vorticity couples with the divergence of the eddy field in the stretching term and the horizontal advection of the background vorticity by the anomalous winds work together to shape the $\beta$-term, and this approximates the tendency of the vorticity anomaly. These indicate that while the description of propagation via the $\beta$-term is tempting and straightforward, other terms in the vorticity budget are required to describe the propagation adequately. 

\begin{figure}[hp]
    \centering
    \includegraphics[width=0.98\linewidth]{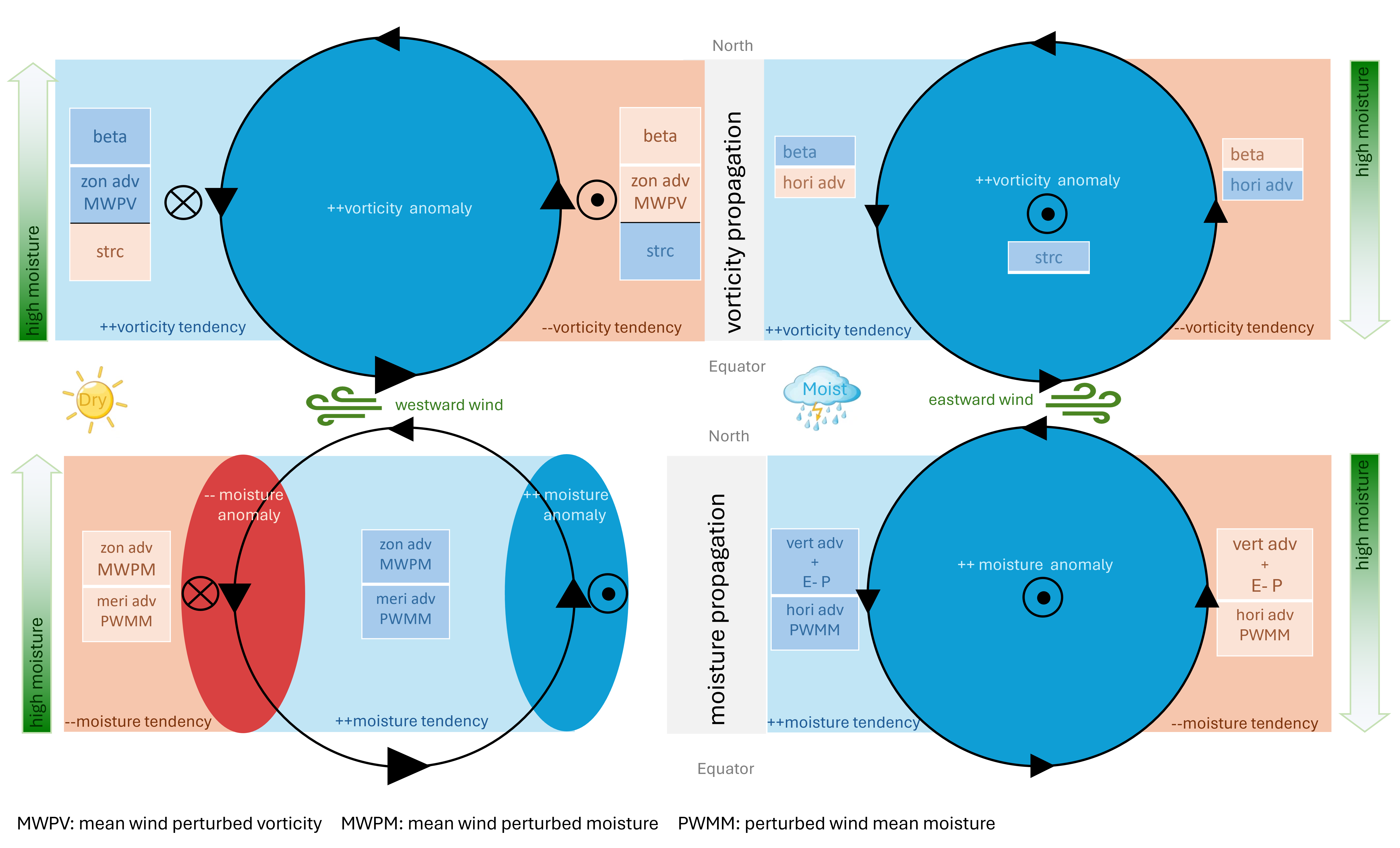}
    \caption{Simplified representation for the evolution of vorticity (upper row) and moisture (lower row) in the lower troposphere, at standard dry (left column) and moist (right column) regions, as mentioned in the paper. The arrows in the circles denote the anomalous cyclonic winds of the gyre. Background wind is indicated by the green wind lines and moisture conditions are shown by arrows filled with gradient in green, where color intensity suggests the moisture amount. Dot and cross denote convergence (upward velocity) and divergence (downward velocity) respectively. Reddish (Bluish) color shading denotes negative (positive) moisture or vorticity quantities with each term written in the picture. This image is not to scale.}
    \label{Sch1}
\end{figure}

The movement of the moisture anomaly is dictated by the mean winds and moisture anomaly gradients or eddy winds and mean moisture gradients. Specifically, in most dry and intermediately moist regions, strong mean easterly winds primarily advect the perturbed moisture westward, and the perturbed meridional wind contributes to the moisture tendency by advecting the mean moisture field, which has a pronounced latitudinal gradient. The moist regions are somewhat diverse in that, in the West Pacific, the advection of perturbed moisture by the mean meridional wind is also crucial for the propagation. Whereas, the Arabian Sea and Bay of Bengal regions are quite distinct in the cause for moisture propagation. Specifically, over the Bay of Bengal and the Arabian Sea, we find an ample accumulation of mean moisture centered around $25^\circ \mathrm{N},95^\circ\mathrm{E}$ (Figure \ref{BG1}), which alters the conventional equator-pole moisture gradient. Additionally, the mean seasonal wind is eastward. The advection of background moisture by the perturbed winds hence encourages the westward propagation of moisture in these two regions. Moreover, this is consistent with the geometry of the eddies as the moisture anomaly is within the gyre on Day 0, and its tendency is highest on the edges of the gyre, as would be expected from the nature of the eddy winds, as they too are largest on the periphery of the QBWO vortex.  Additionally, in all three moist zones (West Pacific, Arabian Sea and Bay of Bengal), the sum of vertical advection, evaporation and precipitation helps to push the moisture westward. Taken together, the coherent evolution of both moisture and vorticity anomalies is described succinctly via the schematic shown in Figure \ref{Sch1}. In summary, our results indicate that the westward propagating boreal summer QBWO is a mean flow driven linear mode in dry and intermediately moist regions, and depends on eddy advection of vorticity and moisture in the very moist regions of the northern hemisphere tropics.

Thus, the propagation of the westward-moving QBWO in the boreal summer is explained from both dynamical and moisture perspectives. The roles of background moisture and its gradients, along with the nature of the prevailing mean flow, are shown to be crucial in determining the structure and propagation of this tropical intraseasonal oscillation. The identification of dominant processes in different moisture regimes provides an immediate application in assessing the fidelity of general circulation models in their representation and development of the QBWO. Moreover, this work offers a pathway toward constructing simplified theoretical models of the QBWO in which moisture and vorticity are explicitly coupled, using the key terms identified here for specific regions of the northern hemisphere tropics. For example, the slow evolution of moisture on QBWO time scales and the prominent role of gyres in advecting background moisture suggest a connection between this mode in the Bay of Bengal and Arabian Sea and the moist quasi-geostrophic framework \citep{joy}. Further explorations, such as the stirring of moisture by QBWO eddies, the coupling of moisture and vorticity through a moist potential vorticity–like framework and relations between vorticity and moisture anomalies in conjunction with divergence and moisture anomalies via weak temperature gradients remain promising avenues for improving our understanding of this mode and its role in the tropical atmosphere. While not the first study to explore QBWO dynamics, our analysis provides additional insight into how moisture-dependent processes shape its evolution, offering useful constraints and guidance for future theoretical development and model evaluation.

\section*{Acknowledgements}
SB acknowledges support from the Prime Minister's Research Fellowship (PMRF), the Government of India. JS acknowledges support from the Indo-Israel joint collaboration (DST/INT/ISR/P-40/2023). BG acknowledges  Anusandhan National Research Foundation (ANRF) research grant SPR/2020/000531.
Both SB and JS would like to acknowledge discussions with Prof. Nili Harnik (Tel Aviv University). SB also acknowledges discussions with Prof. Peter Haynes at the University of Cambridge during his stay as a David Crighton Fellow in the Department of Applied Mathematics and Theoretical Physics. The authors also acknowledge the use of ChatGPT5 for suggestions on improving the language in the manuscript.

\section*{Conflict of interest}
The authors declare no conflict of interest.


\bibliography{sample}

\end{document}